\documentclass[prd,nofootinbib,preprint,superscriptaddress,twocolumn,10pt]{revtex4-1}
\pdfoutput=1
\usepackage{amsmath,amssymb}
\usepackage{epsfig}
\usepackage{graphicx}
\usepackage[usenames,dvipsnames]{color}
\usepackage{subfigure}
\usepackage{slashed}
\usepackage[colorlinks,citecolor=blue]{hyperref}
\usepackage{color}

\usepackage{multirow}

\usepackage{xcolor}
\usepackage{soul}

\def\beq{\begin{equation}}
\def\eeq{\end{equation}}
\def\bea{\begin{eqnarray}}
\def\eea{\end{eqnarray}}
\newcommand{\DNeff}{\Delta N_\text{eff}}
\newcommand{\tdom}{t_\text{dom}}
\newcommand{\tann}{t_\text{ann}}
\newcommand{\tdec}{t_\text{dec}}
\newcommand{\Tdom}{T_\text{dom}}
\newcommand{\Tann}{T_\text{ann}}
\newcommand{\vbl}{v_\text{BL}}
\usepackage[normalem]{ulem}

\begin{document}
\title{Scale of Dirac leptogenesis and left-right symmetry in the light of recent PTA results}
\author{Basabendu Barman}
\email{basabendu88barman@gmail.com}
\affiliation{Institute of Theoretical Physics, Faculty of Physics, University of Warsaw,\\ ul. Pasteura 5, 02-093 Warsaw, Poland}

\author{Debasish Borah}
\email{dborah@iitg.ac.in}
\affiliation{Department of Physics, Indian Institute of Technology Guwahati, Assam 781039, India}

\author{Suruj Jyoti Das }
\email{suruj@iitg.ac.in}
\affiliation{Department of Physics, Indian Institute of Technology Guwahati, Assam 781039, India}

\author{Indrajit Saha}
\email{s.indrajit@iitg.ac.in}
\affiliation{Department of Physics, Indian Institute of Technology Guwahati, Assam 781039, India}
\begin{abstract}
Motivated by the recent release of new results from five different pulsar timing array (PTA) experiments claiming to have found compelling evidence for primordial gravitational waves (GW) at nano-Hz frequencies, we study the consequences for two popular beyond the Standard Model (SM) frameworks, where such nano-Hz GW can arise due to annihilating domain walls (DW). Minimal framework of Dirac leptogenesis, as well as left-right symmetric model (LRSM) can lead to formation of DW due to spontaneous breaking of $Z_2$ symmetry. Considering the NANOGrav 15 yr data, we show that the scale of Dirac leptogenesis should be above $10^7$ GeV for conservative choices of Dirac Yukawa couplings with fine-tuning at the level of the SM. The scale of {\it minimal} LRSM is found to be more constrained $M_{\rm LR} \sim 10^6$ GeV in order to fit the NANOGrav 15 yr data. On the other hand, the {\it non-minimal} LRSM can be compatible with the NANOGrav data for $10^2 \, {\rm TeV} \lesssim M_{\rm LR} \lesssim 10^3$ TeV but with the corresponding $B-L$ breaking scale violating collider bounds.

\end{abstract}
\maketitle

\noindent {\bf Introduction:} Recently, four different pulsar timing array (PTA) experiments namely NANOGrav \cite{NANOGrav:2023gor}, European Pulsar Timing Array (EPTA) together with the first data release from Indian Pulsar Timing Array (InPTA) \cite{Antoniadis:2023ott}, PPTA \cite{Reardon:2023gzh}, all part of the consortium called International Pulsar Timing Array (IPTA) have released their latest findings hinting at a significant evidence for stochastic gravitational waves (GW) background at nano-Hz frequencies. Similar evidence with larger statistical significance has also been reported by the Chinese Pulsar Timing Array (CPTA) collaboration \cite{Xu:2023wog}. While such a signal can be generated by supermassive black hole binary (SMBHB) mergers though with a mild tension, presence of exotic new physics alone or together with SMBHB can make the fit better \cite{NANOGrav:2023hvm}\footnote{Similar conclusions can also be found in~\cite{Antoniadis:2023xlr}.}. Several follow-up papers have also studied the possible origin or implications of this observation from the point of view of dark matter \cite{Ghoshal:2023fhh, Shen:2023pan}, axions or axion-like particles \cite{Yang:2023aak, Guo:2023hyp}, SMBHB \cite{Ellis:2023dgf}, first order phase transition \cite{Megias:2023kiy, Fujikura:2023lkn, Han:2023olf, Zu:2023olm}\footnote{See \cite{Kobakhidze:2017mru} for earlier works.} and associated challenges \cite{Athron:2023mer}, primordial black holes \cite{Franciolini:2023pbf}, primordial magnetic field \cite{Li:2023yaj}, domain walls \cite{Kitajima:2023cek, Bai:2023cqj}, inflation \cite{Vagnozzi:2023lwo, Niu:2023bsr}, cosmic strings \cite{Ellis:2023tsl, Wang:2023len}, scalar induced gravitational waves \cite{NANOGrav:2023hvm} including earlier works \cite{Chen:2019xse}, astrophysical neutrino oscillation \cite{Lambiase:2023pxd} and QCD crossover \cite{Franciolini:2023wjm}. New physics possibilities leading to primordial GW in the nano-Hz regime can also be found in~\cite{Madge:2023cak}.

While GW from domain walls (DW) has already been studied as a possible new physics explanation for PTA results~\cite{NANOGrav:2023hvm, Kitajima:2023cek, Bai:2023cqj}, we consider the consequence for two popular beyond standard model (BSM) scenarios namely, the minimal Dirac leptogenesis and the left-right symmetric model (LRSM). The first model is a type I seesaw realisation for light Dirac neutrino mass with the heavy vector-like neutral fermions being responsible for generating baryogenesis via leptogenesis \cite{Fukugita:1986hr} with light Dirac neutrinos, known as the Dirac leptogenesis scenario \cite{Dick:1999je, Murayama:2002je}. GW probe of high scale leptogenesis models have received considerable attention in recent times. In most of these works \cite{Dror:2019syi, Blasi:2020wpy, Fornal:2020esl, Samanta:2020cdk}, cosmic string (CS) origin of GW has been studied by considering a $U(1)_{B-L}$ framework with in-built heavy Majorana fermions responsible for generating Majorana mass of light neutrinos as well as leptogenesis. The scale of leptogenesis or $U(1)_{B-L}$ breaking scale then decides the amplitude of the CS generated GW spectrum. However, in view of the latest PTA results preferring a positive slope of the GW spectrum, stable CS in such models no longer provide a good fit \cite{NANOGrav:2023hvm}. This raises the prospects for a Dirac leptogenesis model whose minimal version must have a softly broken $Z_2$ symmetry leading to formation of DW followed by generation of GW due to annihilation or collapse. While a general study related to GW probe of minimal Dirac leptogenesis was carried out in \cite{Barman:2022yos} (and subsequently in \cite{King:2023cgv} for Dirac seesaw), here we consider the implications of recent PTA findings on the scale of Dirac leptogenesis. On the other hand, GW probe of LRSM considering DW as the source have been studied in earlier works \cite{Craig:2020bnv,Borah:2022wdy}. DW arise due to spontaneous breaking of parity in such models. While earlier works considered the detection aspects of this model, we now constrain the scale of left-right symmetry considering the latest PTA data. While both the models can explain the latest PTA data, the allowed parameter space remains squeezed to a tiny window, which should face more scrutiny with future data.

\vspace{0.5cm}
\noindent
{\bf Domain walls as source of GW:}
\label{sec:GW}
Domain wall is a two-dimensional topological defect arising from spontaneous breaking of discrete symmetries  \cite{Zeldovich:1974uw, Kibble:1976sj, Vilenkin:1981zs}. With the expansion of the universe, the energy density of DW falls slower compared to that of radiation or ordinary matter, having the potential to start dominating the energy density of the universe and ruin the successful predictions of standard cosmology. Such a disastrous situation can be prevented if DW are made unstable or diluted or if they have asymmetric initial field fluctuations ~\cite{Coulson:1995nv, Krajewski:2021jje}.

In minimal model of Dirac leptogenesis~\cite{Barman:2022yos} as well as left-right symmetric models \cite{Borah:2022wdy}, such DW arises due to the spontaneous breaking of a $Z_2$ symmetry. If we consider a $Z_2$-symmetric potential of a scalar field $\varphi$, it is straightforward to show the existence of two different vacua $\langle \varphi \rangle = \pm u$. It is also possible to find a static solution of the equation of motion given the two vacua to be realized at $x \to \pm \infty$,
\begin{align}
	\varphi({\bf x}) = u\,\tanh\left( \sqrt{\frac{\lambda_\varphi}{2}}\,u\,x \right)\,,
\end{align}
representing a domain wall extended along the $x = 0$ plane. Here $\lambda_\varphi$ is the quartic self-coupling of the scalar field. The DW width is $\delta \sim m_\varphi^{-1} = (\sqrt{2\lambda_\varphi}\,u)^{-1}$. Another key parameter, known as the DW tension is given by
\begin{align}
	\sigma_w = \int_{-\infty}^{\infty} dx \,\rho_\varphi = \frac{2\sqrt 2}{3}\,\sqrt{\lambda_\varphi}\,u^3 = \frac{2}{3}\,m_\varphi\,u^2 \sim u^3\,,
\end{align}
where $\rho_\varphi$ denotes (static) energy density of $\varphi$ and in the last step, $m_\varphi \sim u$ is used.

Assuming the walls to be formed after inflation, the simplest way to make them disappear is to introduce a small pressure difference \cite{Zeldovich:1974uw, Vilenkin:1981zs, Sikivie:1982qv, Gelmini:1988sf, Larsson:1996sp}, a manifestation of a soft $Z_2$-breaking term. Such a pressure difference or equivalently, a bias term in the potential $\Delta V$ needs to be sufficiently large to ensure DW disappearance prior to the epoch of big bang nucleosynthesis (BBN) that is, $t_{\rm BBN} > t_{\rm dec} \approx \sigma_w/\Delta V$. It is also important to take care of the fact that the DW disappear before dominating the universe, requiring $t_\text{dec}<t_\text{dom}$, where $t_\text{dom}\sim M_P^2/\sigma_w$ and $M_P$ is the {\it reduced} Planck mass. Both of these criteria lead to a lower bound on the bias term $\Delta V$. However, $\Delta V$ can not be arbitrarily large as it would otherwise prevent the percolation of both the vacua separated by DW. Such decaying DW therefore can emit GW~\cite{Kadota:2015dza, Hiramatsu:2013qaa, Krajewski:2016vbr, Nakayama:2016gxi, Dunsky:2020dhn, Babichev:2021uvl, Ferreira:2022zzo, Deng:2020dnf, Gelmini:2020bqg, Saikawa:2017hiv}. { At peak frequency $f_{\rm peak}$, the spectral energy density can be estimated as}~\cite{Kadota:2015dza, Hiramatsu:2013qaa}
\begin{align}
    \Omega_{\rm GW}h^2\,(t_0) \Big|_{\rm peak} & \simeq 5.2 \times 10^{-20}\,\tilde{\epsilon}_{\rm gw}\,A_w^4 \left (\frac{10.75}{g_*}\right)^{1/3} \nonumber \\
    & \times\left (\frac{\sigma_w}{1 \, {\rm TeV}^3 }\right)^4 \,\left (\frac{1 \, {\rm MeV}^4}{\Delta V}\right)^2\,,
\end{align}
with $t_0$ being the present time. Away from the peak, the amplitude varies as\footnote{ The low-frequency spectrum of GW from melting DWs~\cite{Babichev:2021uvl,Babichev:2023pbf} characterized by a {\it time-dependent} tension, contrary to the constant tension DWs as discussed here, behaves as $f^2$, without violating the causality.}
\begin{align}
	\Omega_{\rm GW} \simeq \Omega_{\rm GW}\Big|_{\rm peak} \times 
	\begin{cases}
		\displaystyle{\left( \frac{f_{\rm peak}}{f} \right)} & {\rm for}~~f>f_{\rm peak}\\
		&\\
		\displaystyle{\left( \frac{f}{f_{\rm peak}} \right)^3 }& {\rm for}~~f<f_{\rm peak}
	\end{cases}\,,\label{eqn:omgw1}
\end{align}
where the peak frequency is given by
\begin{align}
    f_{\rm peak} (t_0) & \simeq \textcolor{blue}{4} \times 10^{-9} \, {\rm Hz}\, A^{-1/2} \nonumber \\
    & \times \left ( \frac{ 1\, {\rm TeV}^3}{\sigma_w} \right)^{1/2}\, \left ( \frac{\Delta V}{1\, {\rm MeV}^4} \right)^{1/2}\,.
\end{align}
In the above expressions, $A_w$ is the area parameter~\cite{Caprini:2017vnn, Paul:2020wbz} $\simeq 0.8$ for DW arising from $Z_2$ breaking, and $\tilde{\epsilon}_{\rm gw}$ is the efficiency parameter $\simeq$ 0.7~\cite{Hiramatsu:2013qaa}. Note that the above spectrum can be obtained from a general parametrisation $S(f/f_{\rm peak})$
\begin{equation}
    S(x) = \frac{(a+b)^c}{(bx^{-a/c}+ax^{b/c})^c}
\end{equation}
for $a=3$ (required by causality~\cite{Cai:2019cdl,Hook:2020phx}) and $b \approx c \approx 1$ (suggested by simulation~\cite{Hiramatsu:2013qaa}). Utilising these values of $a,b$ and $c$, togeteher with $x\gg1$ (or $x\ll 1$), we can produce Eq.~\eqref{eqn:omgw1}. However, as noted in \cite{NANOGrav:2023hvm, Ferreira:2022zzo}, the values of $b,c$ may depend upon the specific DW annihilation mechanism or regime all of which have not been explored in numerical simulations yet. This allows one to vary $b,c$ to get a better fit with the PTA data \cite{NANOGrav:2023hvm}.

When the GW production is ceased after the annihilation of the domain walls, the energy density of GW redshifts mimicking that of the SM radiation. As a result, GW itself acts as an additional source of radiation with the potential to alter the prediction of BBN. Thus, an excess of the GW energy density around $T \lesssim\mathcal{O}(\text{MeV})$, can be restricted by considering the limits on the number of relativistic degrees of freedom  from CMB and BBN, encoded in
$\DNeff$. This, in turns, puts a bound on the amplitude of GW spectrum, demanding~\cite{Maggiore:1999vm,Caprini:2018mtu} 
\begin{align}\label{eq:gw-neff}
\Omega_\text{GW}\,h^2\lesssim 5.6\times 10^{-6}\,\DNeff\,.    
\end{align}
Here we consider several projected limits on $\DNeff$ on top of the existing limit from Planck: $\Delta N_\text{eff} \lesssim 0.28$ at 95\% CL~\cite{Planck:2018vyg}. This bound is shown by the solid gray horizontal line in Fig.~\ref{fig:fig2}. Once the baryon acoustic oscillation (BAO) data are included, the measurement becomes more stringent: $N_\text{eff} = 2.99 \pm 0.17$. A combined BBN+CMB analysis shows $N_\text{eff} = 2.880 \pm 0.144$, as computed in Ref.~\cite{Yeh:2022heq}. This constraint is denoted by the dashed horizontal line. On the other hand, upcoming CMB experiments like CMB-S4~\cite{Abazajian:2019eic} and CMB-HD~\cite{CMB-HD:2022bsz} will be able to probe $\DNeff$ as small as $\sim 0.06$ and $\sim 0.027$, respectively. These are indicated by dot-dashed and dotted lines respectively. The next generation of satellite missions, such as COrE~\cite{COrE:2011bfs} and Euclid~\cite{EUCLID:2011zbd}, leads to $\DNeff \lesssim 0.013$, as shown by the large dashed line. 

It should be noted that we are ignoring the friction effects between the domain walls and the thermal plasma \cite{Nakayama:2016gxi, Galtsov:2017udh, Blasi:2022ayo}. Such friction effects can be important when the field responsible for symmetry breaking or constituting the wall has large couplings with the SM bath particles, leading to smaller GW amplitude than that without friction discussed here. Since the scalar fields responsible for symmetry breaking has tiny couplings with the SM plasma in the models discussed here, such effects can be ignored \cite{Babichev:2021uvl}.

\begin{figure}[htb!]
\includegraphics[scale=0.35]{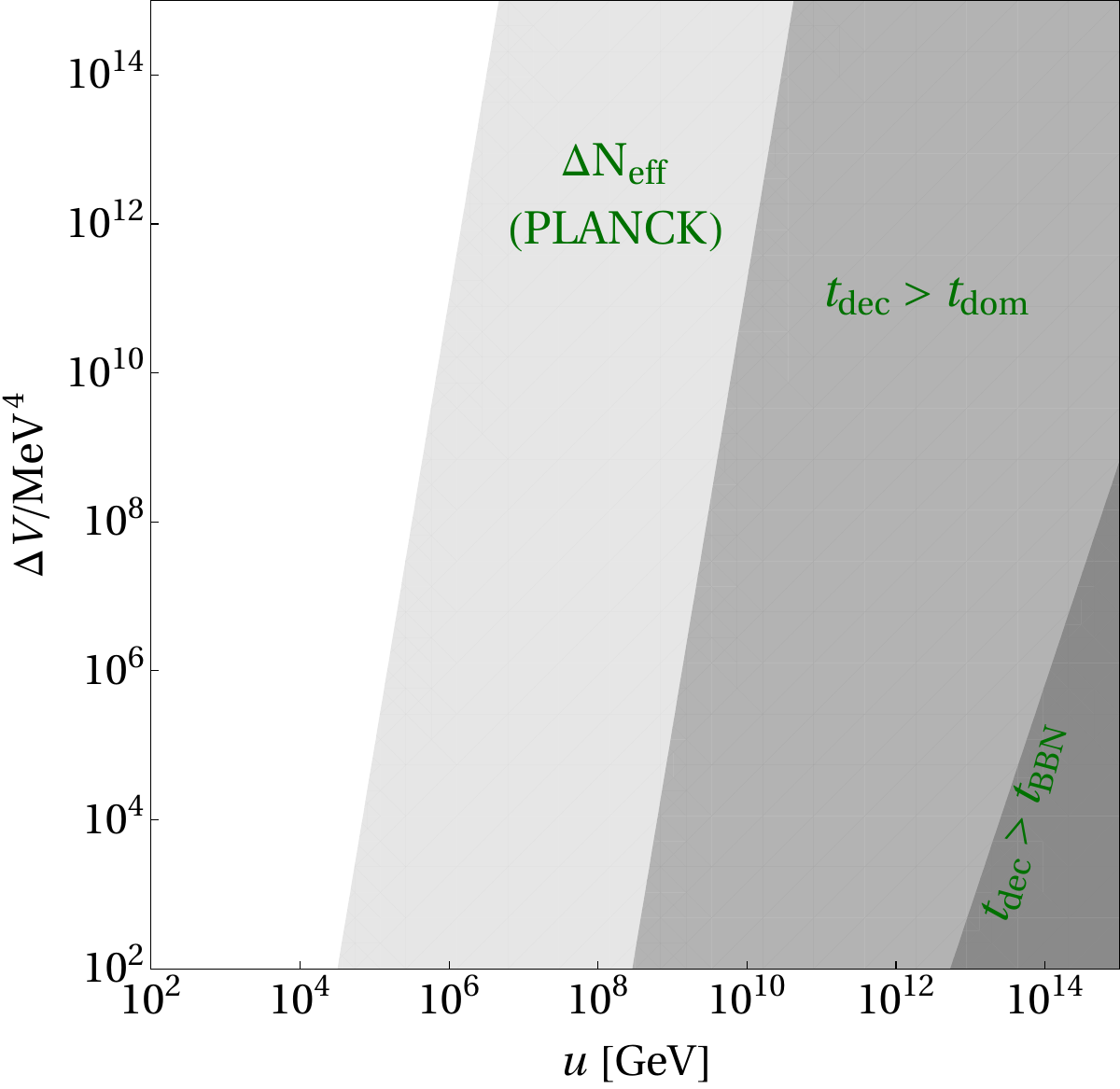}
\caption{Bound on the size of VEV and the bias term. All shaded regions are disallowed (see text for details).}
\label{fig:fig1} 
\end{figure}

\begin{figure}[htb!]
\includegraphics[scale=0.4]{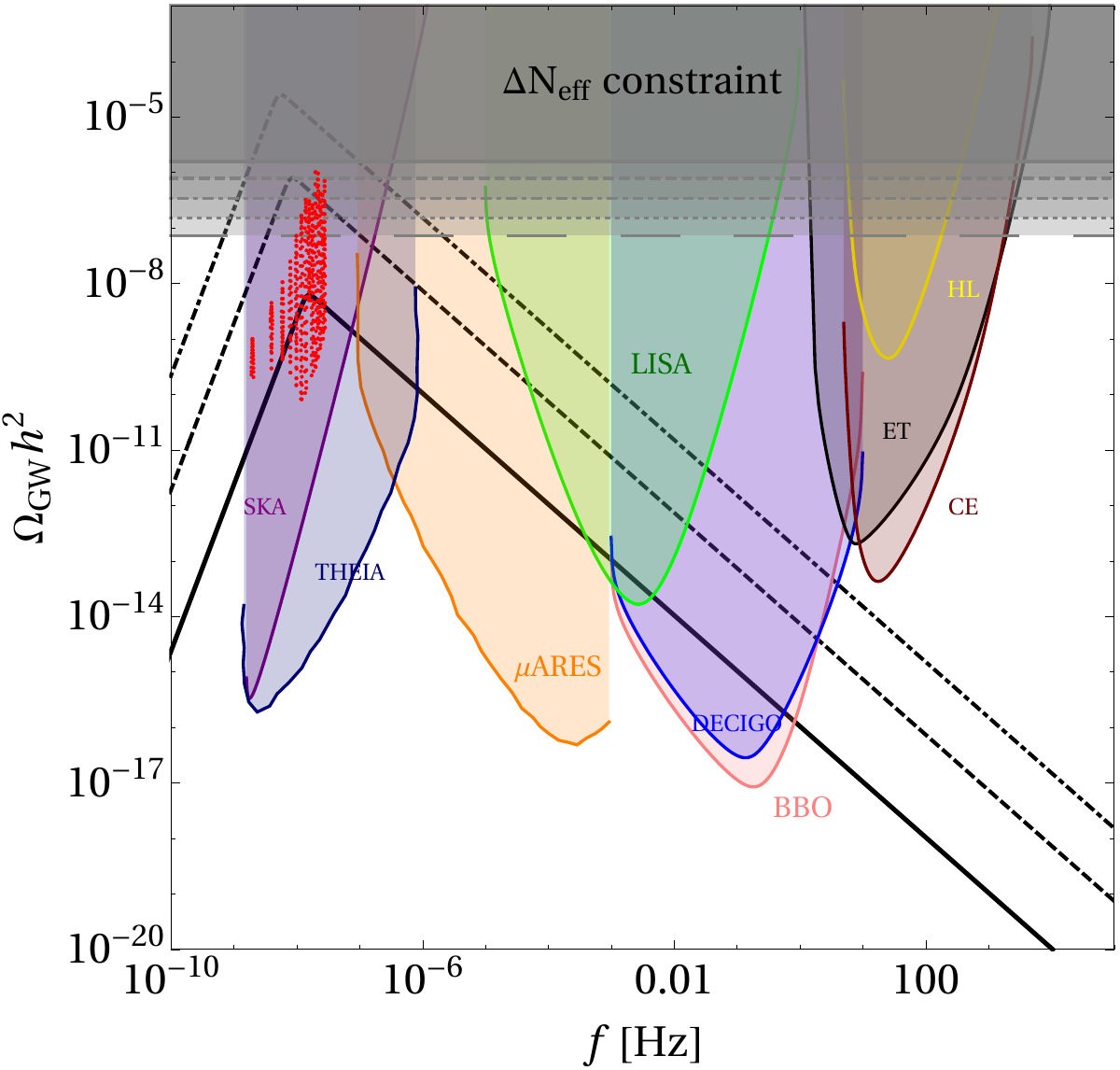}
\caption{Spectrum of gravitational wave from DW decay, where we show sensitivities of several GW experiments. The black curves correspond to the chosen benchmark points for $a=3\,,b=1\,,c=0.3$. The gray region marked as ``$\DNeff$ constraint" is disallowed from overproduction of GW energy density (see text for details). The red points correspond to the NanoGrav 15 yr observation~\cite{NANOGrav:2023gor}.}
\label{fig:fig2} 
\end{figure}

In Fig.~\ref{fig:fig1} we summarize bounds on the VEV $u$ and the bias term $\Delta V$, where all the shaded regions are disallowed from (i) decay of the DWs post BBN (darker gray), where $t_\text{dec}>1$ sec, (ii) DW domination (lighter gray) or $t_\text{dom}<t_\text{dec}$ and (iii) $\DNeff$ bound from PLANCK on excessive GW energy density (light gray). This leaves us with the white region in-between that is allowed, from where we choose our benchmark points (BP), as indicated in Tab.~\ref{tab:BP}. It is important to note here, we also consider $\Delta V\ll u^4$ to prevent the percolation of both the vacua separated by DW. However, such a condition is trivially satisfied in the regime we are interested in.

\begin{table}
    \centering
    \begin{tabular}{|c|c|c|c|c|}
    \hline
     & $u$ (GeV) & $\Delta V (\text{MeV})^4$ \\
    \hline  
    {\tt BP1} & $2\times 10^5$ & $10^8$ \\
    {\tt BP2} & $3\times 10^5$ & $10^8$ \\
    {\tt BP3} & $4\times 10^5$ & $10^8$ \\
    \hline
    \end{tabular}
    \caption{Details of the benchmark points (BPs) used in Fig.~\ref{fig:fig2} and Fig.~\ref{fig:fig3}.}
  \label{tab:BP}
\end{table}

The GW spectrum corresponding to the BPs in Tab.~\ref{tab:BP} is illustrated in Fig.~\ref{fig:fig2}. As explained before, we distinctively see a blue-tilted pattern for $f<f_\text{peak}$, while the spectrum is red-tilted in the opposite limit. Here we project limits from BBO~\cite{Yagi:2011wg}, LISA~\cite{AmaroSeoane2012LaserIS}, DECIGO~\cite{Kawamura:2006up}, ET~\cite{Punturo_2010}, CE~\cite{LIGOScientific:2016wof}, THEIA~\cite{Garcia-Bellido:2021zgu}, HL (aLIGO)~ \cite{LIGOScientific:2014pky}, $\mu$ARES~\cite{Sesana:2019vho} and SKA~\cite{Weltman:2018zrl}. In this plot, the range of GW spectrum from NANOGrav results~\cite{NANOGrav:2023gor} is shown by the red points. The gray-shaded region is completely disallowed from $\DNeff$ bound on overproduction of $\Omega_\text{GW}$ as discussed before, depending on the sensitivity of a particular experiment. As one can already notice, BP3 is already ruled out from PLANCK bound, while BP1 is beyond the reach of any future experiments proposed so far.

\begin{figure}[htb!]
\includegraphics[scale=0.4]{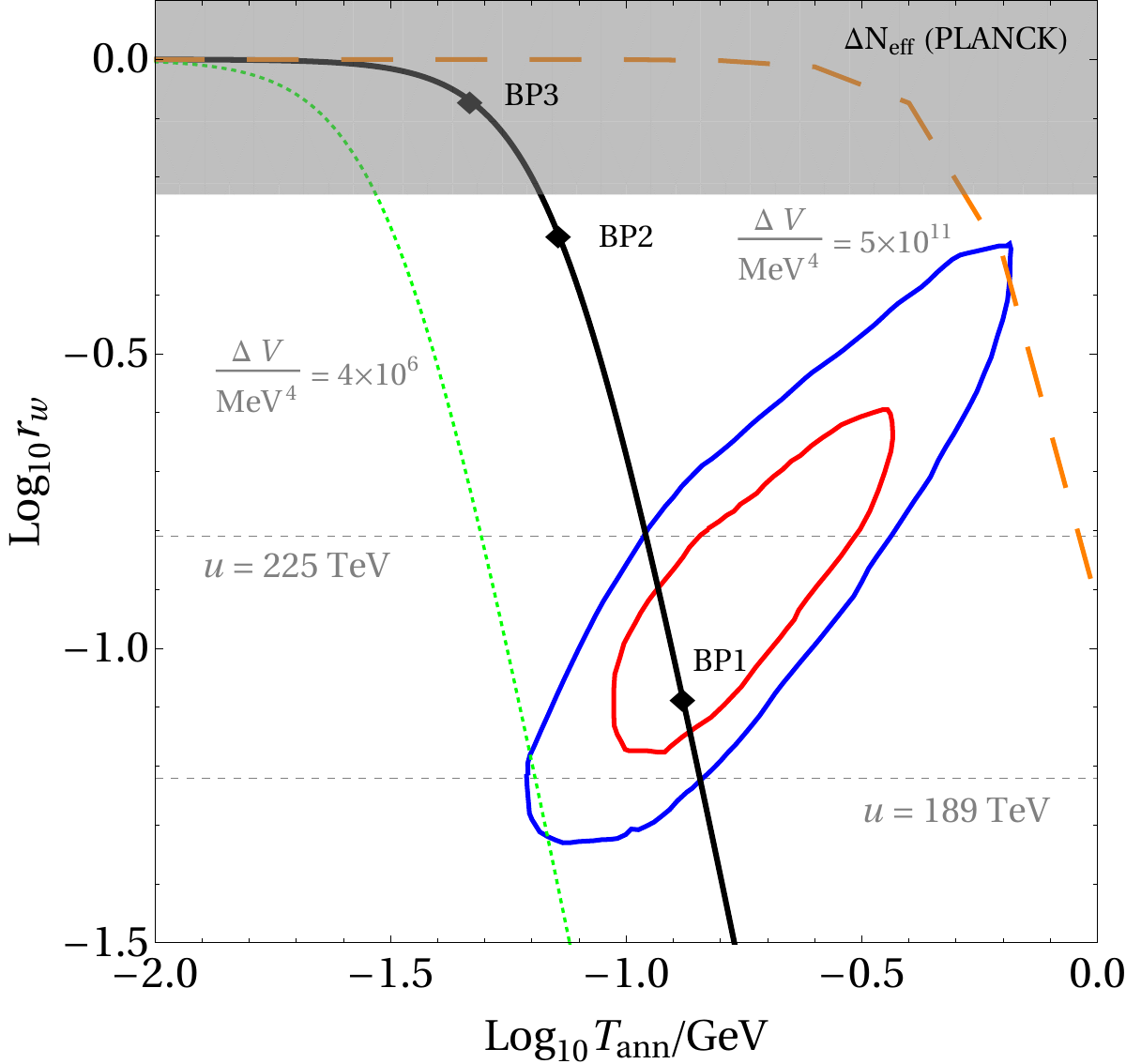}
\caption{Viable parameter space in the bi-dimensional plane of $\Tann$-$r_w$, based on the NanoGrav 15 yr dataset~\cite{NANOGrav:2023gor}. The black solid contour corresponds to the $\Delta V$-value relevant for the BPs in Tab.~\ref{tab:BP}. The red and blue contours correspond to 1- and 2$\sigma$ CL respectively. The gray shaded region is disallowed from $\DNeff$ bound from Planck.}
\label{fig:fig3} 
\end{figure} 

Corresponding to the two epochs $\tdom$ and $\tann$ stated before, we define two temperatures $\Tdom$ and $\Tann$. We are typically interested in the regime $\Tann>\Tdom$, i.e., the DWs disappear before they dominant the Universe. Following~\cite{Bai:2023cqj}, $\Tann$ reads
\begin{align}\label{eq:Tann}
& \Tann\simeq 120\,\text{MeV}\,\sqrt{\frac{\Delta V/\text{MeV}^4}{10^8}}\, \left(\frac{A_w}{0.8}\right)^{-1/2}
\nonumber\\&
\left(\frac{\sigma_w/\text{GeV}^3}{10^{16}}\right)^{-1/2}\,\left(\frac{g_*(\Tann)}{10}\right)^{-1/4}\,,
\end{align}
implying, for larger surface tension, it takes longer for the walls to collapse, while for larger bias the opposite happens. We also define another quantity
\begin{align}\label{eq:rw}
& r_w = \frac{\rho_r}{1+\rho_r}\,,    
\end{align}
where
\begin{align}
& \rho_r =  \frac{\rho_w(\Tann)}{\rho_R(\Tann)}\simeq 0.14\,\left(\frac{A_w}{0.8}\right)^2\,\left(\frac{\sigma_w/\text{GeV}^3}{10^{16}}\right)^2
\nonumber\\&
\left(\frac{10^8\,\text{MeV}^4}{\Delta V}\right)\,, 
\end{align}
which quantifies the energy density contained within the DW compared to that of radiation. 

We show the compatibility of our relevant model parameters, namely, the bias term $\Delta V$ and the VEV $u$ (equivalently, the strain $\sigma$)
in Fig.~\ref{fig:fig3} with the NANOGrav data, utilizing Eq.~\eqref{eq:Tann} and Eq.~\eqref{eq:rw}. We superimpose the 1 and 2$\sigma$ contours (shown by red and blue solid curves) provided by the NANOGrav  result~\cite{NANOGrav:2023gor}. As one can see, {\tt BP1} lies well within the $1\sigma$ contour, while the other two BPs are well off. For {\tt BP1}, $a=3\,,b\in[0.5\,,1]\,,c\in[0.3\,,3]$ is needed to be in compliance within 1$\sigma$ of NANOGrav data for frequency $f\in [2\times 10^{-9}\,,f_{\rm yr}]$ Hz, where $f_{\rm yr}=1\,\text{yr}^{-1}\approx 3\times 10^{-8}$ Hz. It is possible to derive a lower and upper bound on the VEV for a fixed $\Delta V$, as denoted by the gray dashed horizontal lines for $\Delta V=10^8\,\text{MeV}^4$. Thus, for $\Delta V=10^8\,\text{MeV}^4$, we find $189\lesssim u/\text{TeV}\lesssim 225$, in compliance with NANOGrav $2\sigma$ contour. On the other hand, the viable range of bias term, that lies within $2\sigma$ CL of NANOGrav result, turns out to be $4\times 10^6\lesssim\Delta V/\text{MeV}^4\lesssim5\times 10^{11}$, as shown by the green dotted and orange dashed curves. Depending on the choice of $\Delta V$, the upper limit on $u$ can be pushed to larger values. One can also project the $\DNeff$ bound on the same plane (by trading $\{\sigma\,,\Delta V\}$ with $\{\Tann\,,r_w\}$) ruling out the region of the parameter space that results in overproduction of GW [cf. Eq.~\eqref{eq:gw-neff}]. This is shown by the gray shaded region, where we have used the 2-$\sigma$ bound from Planck. This rules out {\tt BP3}, as already seen in Fig.~\ref{fig:fig1}.
Note that, the limits obtained on VEV $u$ together with that on the bias $\Delta V$, satisfy $\tdom<t_\text{dec}$ and $\tdec<t_\text{BBN}$, obeying $\Delta V\ll u^4$. 

\vspace{0.5cm}
\noindent
{\bf Consequence for Dirac leptogenesis:}
\label{sec:dir-lep}
In the minimal model of Dirac leptogenesis or Dirac neutrino seesaw \cite{Barman:2022yos}, the standard model (SM) is extended by three copies of vector-like neutral singlet fermions $N_{L,R}$ and three copies of right chiral part $(\nu_R)$ of light Dirac neutrinos. A real singlet scalar field $\varphi$ is introduced to couple $\nu_R$ with $N$. A $Z_2$ symmetry under which $\varphi, \nu_R$ are odd, prevents direct coupling of the SM lepton doublet $L$ with $\nu_R$ via SM Higgs $H$. The relevant part of the Yukawa Lagrangian can be written as
\begin{align}
    -\mathcal{L}_Y \supset Y_L\, \overline{L}\,\widetilde{H}\,N_R + M_N\, \overline{N}\,N + Y_R\,\overline{N_L}\,\varphi \,\nu_R + {\rm h.c.}
\end{align}
After the neutral components of $H$ and $\varphi$ acquire VEV $v,\, u$ respectively, light Dirac neutrino mass arises from the Type-I seesaw equivalent for Dirac neutrino as 
\begin{equation}
    m_{\nu} = \frac{1}{\sqrt{2}}Y_L\,M^{-1}_N\,Y_R\,v\,u\,
\end{equation}
with $M_N$ being the scale of Dirac seesaw. The same heavy fermions $N_{L,R}$ can have out-of-equilibrium decay to achieve successful Dirac leptogenesis. Although no net lepton asymmetry is produced due to total lepton number conservation, it is possible to create equal and opposite lepton asymmetries in left and right sectors due to CP violating out-of-equilibrium decays $N \rightarrow L\,H$ and $N \rightarrow \nu_R\,\varphi$ respectively. The $CP$ asymmetry parameter is given as \cite{Cerdeno:2006ha}
\begin{align}
    \epsilon \simeq -\frac{1}{8\pi}\frac{M_1}{uv}\frac{{\rm  Im}[(Y_R m^{\dagger}_{\nu}Y_L)_{11}]}{(Y_RY^{\dagger}_R) + (Y_LY^{\dagger}_L)},
    \label{eq:cp}
\end{align}
where $v = 246$ GeV and $M_1$ is the lightest heavy fermion mass. If a net lepton asymmetry is generated before the sphaleron decoupling epoch, it is possible to create a net baryon asymmetry. However, in order to prevent the left sector lepton asymmetry from being washed out, it is important to prevent the equilibration of left and right sectors, leading to a condition 
\begin{align}
\Gamma_{L-R} &\sim \frac{|Y_L|^2\,|Y_R|^2}{M^2_1}\,T^3 < \mathcal{H}(T)\,,
\end{align}
where $\mathcal{H}(T)=\frac{\pi}{3}\,\sqrt{\frac{g_\star}{10}}\,\frac{T^2}{M_P}$ is the Hubble parameter. We do not compute the lepton asymmetry here and refer the reader to earlier works \cite{Cerdeno:2006ha, Borah:2016zbd, Barman:2022yos} where explicit Boltzmann equations were solved, and corresponding parameter space has been obtained. Depending upon the scale of leptogenesis $M_1$, one can consider either quasi-degenerate or hierarchical heavy fermions to achieve the desired CP asymmetry while being consistent with sub-eV Dirac neutrino mass.

As the $Z_2$-odd scalar $ \varphi $ acquires a non-zero VEV $u$, it leads to the formation of domain walls\footnote{ Because of the non-zero VEV, $\varphi$ can mix with the SM Higgs doublet $(H)$ via a portal interaction $\lambda_p\,|\varphi|^2\,|H|^2$, that leads to its decay into the SM. One can always tune the mixing parameter such that $\varphi$ decays efficiently before the onset of BBN.}. One can introduce a bias term $\Delta V$ (which breaks $Z_2$ symmetry softly) in the scalar potential which eventually lead to the disappearance of domain walls. Now, for $u \gtrsim 190$ TeV, preferred from NANOGrav 2023 data as discussed before, and considering light neutrino mass $m_\nu \leq  0.1$ eV, we get $M_N > y^2 \times 10^{17}$ GeV. This implies that, for order one Yukawa couplings, the scale of Dirac leptogenesis is even above the upper limit on reheating temperature, disfavouring the possibility of thermal Dirac leptogenesis. If the Yukawa couplings are made as low as electron Yukawa coupling, we have $M_N > 10^7$ GeV, keeping it at intermediate scale. To summarize, the possibility of low scale { $Z_2$-symmetric} Dirac leptogenesis is disfavoured unless we tune Yukawa couplings involved in Dirac seesaw more than what we have in the SM. 

It is worth stressing that Dirac leptogenesis can be realised without $Z_2$ symmetry too and in those setups there will not be any DW formation. One simple alternative is to consider a $U(1)$ global symmetry under which $\phi$, now a complex scalar, and $\nu_R$ transform non-trivially. Soft breaking of such global $U(1)$ symmetry, required to generate light Dirac neutrino mass, leads to a light pseudo-Goldstone boson with interesting phenomenological consequences. On the other hand, for gauged $U(1)$ symmetry, an additional massive gauge boson will arise in the spectrum. If such additional neutral bosons couple to the heavy fermions $(N)$ responsible for generating lepton asymmetry, they can lead to dilution of asymmetry while keeping $N$ in equilibrium for longer epochs \cite{Mahanta:2021plx}. As far as topological defects are concerned, such models with $U(1)$ symmetries can lead to cosmic strings which can have their own GW signatures. Since the number of new degrees of freedom in $Z_2$-symmetric Dirac leptogenesis is less than scenarios with $U(1)$ or higher symmetries, we have referred to it as the minimal Dirac leptogenesis.

\vspace{0.5cm}
\noindent
{\bf Consequence for left-right symmetry:} Left-right symmetric models \cite{Pati:1974yy, Mohapatra:1974hk, Mohapatra:1974gc, Senjanovic:1975rk, Senjanovic:1978ev, Mohapatra:1977mj, Mohapatra:1980qe, Mohapatra:1980yp, Lim:1981kv, Gunion:1989in,Deshpande:1990ip,Duka:1999uc, FileviezPerez:2008sr} have been one of the most well-studied BSM frameworks where the SM gauge symmetry is extended to $ \rm SU(3)_c\times SU(2)_L\times SU(2)_R\times U(1)_{B-L}$. In addition to the gauge symmetry, the model also has an in-built $Z_2$ symmetry or parity symmetry ($\mathbb{P}$) under which the left and right sector fields are interchanged, keeping the framework parity or left-right symmetric. While a detailed discussion related to GW probe of LRSM can be found in \cite{Craig:2020bnv,Borah:2022wdy, Borboruah:2022eex}, here we briefly summarise the implications of 2023 PTA data on the scale of left-right symmetry. 
\begin{figure}[htb!]
\includegraphics[scale=0.4]{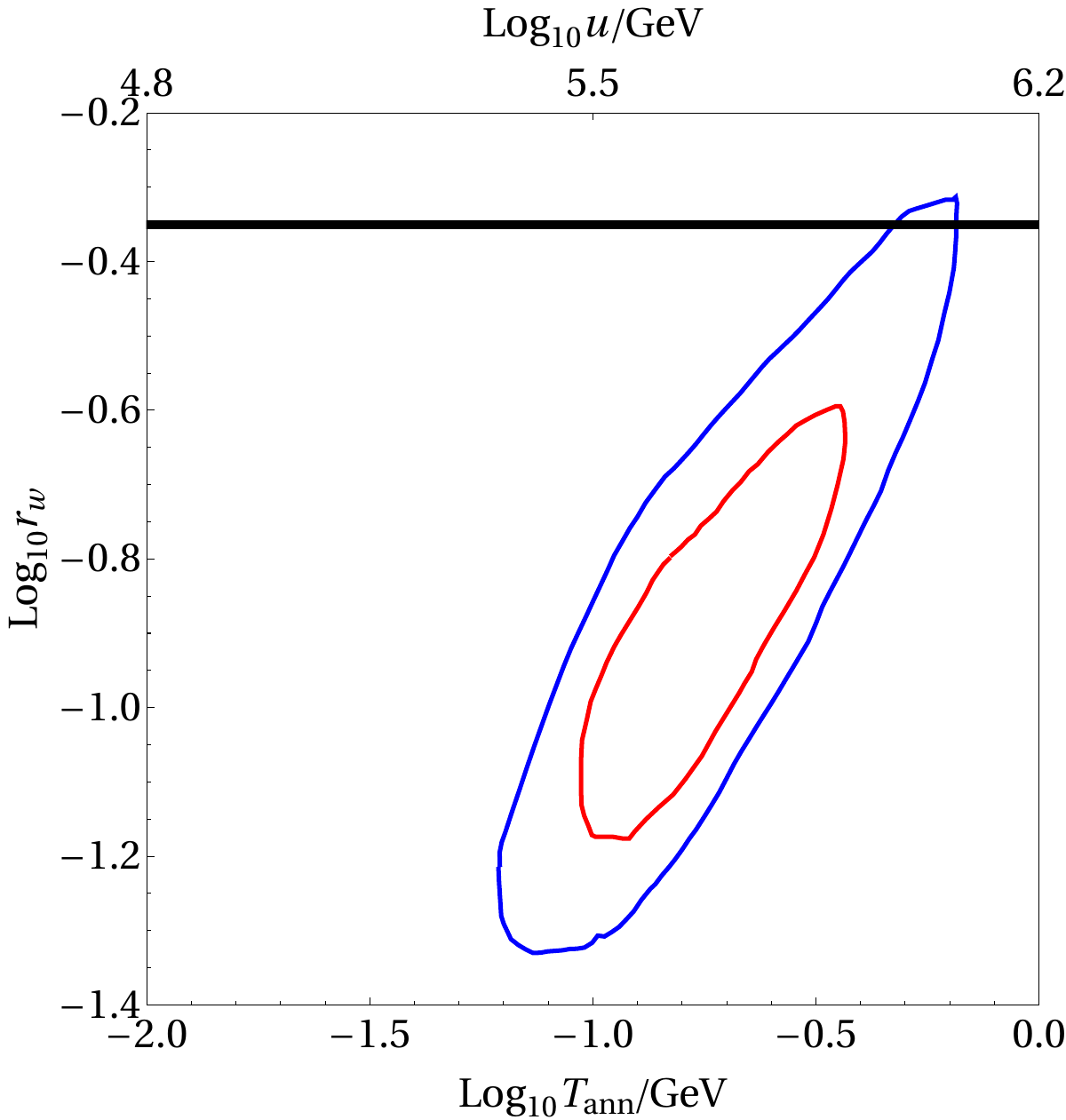}
\caption{Same as Fig.~\ref{fig:fig3}, but for the minimal LRSM. Along the black solid line the left-right symmetry breaking scale $u$ is varying.}
\label{fig:fig4} 
\end{figure} 
\begin{figure}[htb!]
\includegraphics[scale=0.4]{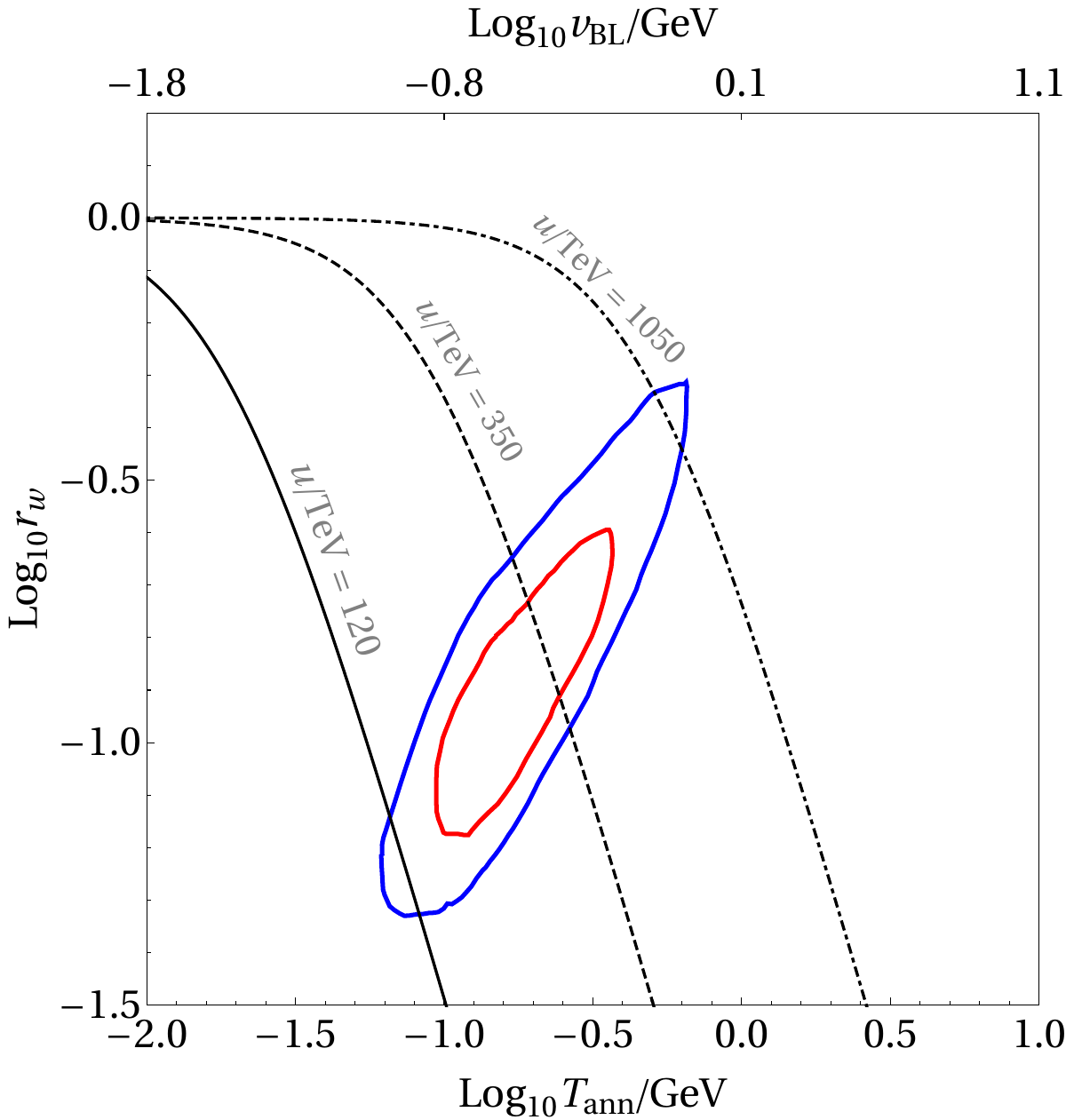}
\caption{Same as Fig.~\ref{fig:fig3}, but for the non-minimal LRSM. Different black curves correspond to different choices of left-right symmetry breaking scale $u$, as denoted.}
\label{fig:fig5} 
\end{figure} 

Unlike in the $Z_2$-odd scalar singlet model discussed above, the LRSM gauge symmetry does not allow arbitrary bias terms. It is possible to generate such bias term via higher dimensional operators invariant under gauge symmetry but explicitly breaking the parity symmetry. While Planck scale effects are expected to break any global symmetries like parity \cite{Abbott:1989jw, Kallosh:1995hi, Hawking:1975vcx}, the corresponding bias term can lead to DW disappearance \cite{Rai:1992xw,Lew:1993yt}. The minimal LRSM has three different types of scalars namely, $\Phi \equiv (1,2,2,0), \Delta_L \equiv (1,3,1,2), \Delta_R \equiv (1,1,3,2)$ where the numbers in the brackets are the quantum numbers corresponding to the LRSM gauge group $SU(3)_c \times SU(2)_L \times SU(2)_R \times U(1)_{B-L}$. Left and right handed fermions transform as doublets under $SU(2)_L, SU(2)_R$ respectively. Quark and lepton fields are represented as $Q_L \equiv (3,2,1,1/3), Q_R \equiv (3,1,2,1/3), \ell_L \equiv (1,2,1,-1), \ell_R \equiv (1,1,2,-1)$. Under parity $\mathbb{P}$, left and right sector fields get interchanged as
$$ Q_L \leftrightarrow Q_R, \ell_L \leftrightarrow \ell_R, \Delta_L \leftrightarrow \Delta_R, \Phi \leftrightarrow \Phi^{\dagger}. $$
This not only ensures the equality of left and right sector gauge couplings $g_L = g_R$, but also relates the Yukawa and scalar potential couplings of these two sectors.

The neutral component of the scalar triplet $\Delta_R$ acquires a non-zero VEV breaking $SU(2)_R \times U(1)_{B-L} \times \mathbb{P}$ into $U(1)_Y$ of the SM. At a later stage, the electroweak gauge symmetry gets spontaneously broken to $U(1)_{\rm em}$ by the neutral components of scalar bidoublet $\Phi$. Thus, the symmetry breaking pattern is 
\begin{align}
 SU(2)_L \times SU(2)_R \times U(1)_{B-L} \times \mathcal{P} \quad \underrightarrow{\langle
\Delta_R \rangle} \nonumber \\
\quad SU(2)_L\times U(1)_Y \quad \underrightarrow{\langle \Phi \rangle} \quad U(1)_{\rm em}.
\end{align}
While this is the desired symmetry breaking pattern, it is equally probable for left sector scalar field $\Delta_L$ to acquire non-zero VEV. This leads to left and right sector vacua separated by domain walls. It is also possible to replace the pair of triplets $\Delta_{L,R}$ by a pair of doublets $H_{L,R}$ while achieving the same symmetry breaking pattern. In either of these minimal models, the bias term or soft $\mathbb{P}$ breaking term can arise from dimension six operators given by
\begin{equation}
V_{\text{NR}} \supset f_L \frac{(\Sigma^{\dagger}_L \Sigma_L )^3}{M^2_{\rm P}} + f_R \frac{(\Sigma^{\dagger}_R \Sigma_R )^3}{M^2_{\rm P}}
\end{equation}
where $\Sigma_{L,R} \equiv \Delta_{L,R}, H_{L,R}$ depending upon the type of LRSM. This leads to a bias term in the minimal model given by $\Delta V \sim u^6/M_P^2$, where $u$ is the $SU(2)_R \times U(1)_{B-L}$ as well as parity breaking scale. Due to the dependence of the bias term on the scale of left-right breaking, the constraint on left-right symmetry breaking scale is stronger than what we had on the scale of Dirac leptogenesis. As shown in Fig. \ref{fig:fig4}, the scale of left-right symmetry should be approximately around $\sim 10^6$ GeV (shown by the black solid horizontal line), in order to be in agreement with NANOGrav 15 yr data at $2\sigma$.

Similarly, one can also check the status of non-minimal LRSM frameworks in the light of the recent PTA results. Unlike the minimal LRSM, in the non-minimal scenario, the symmetry of LRSM is broken down to that of the SM in more than one steps. For illustrative purpose, we consider a two-step symmetry breaking chain leading to
\begin{align}
& SU(2)_L\otimes SU(2)_R\otimes U(1)_{B-L}\otimes \mathbb{P} \xrightarrow[]{u} 
\nonumber\\&
SU(2)_L\otimes U(1)_R\otimes U(1)_{B-L} \xrightarrow[]{\vbl}
\nonumber\\&
SU(2)_L\otimes U(1)_Y\xrightarrow[]{v_\text{EW}} U(1)_\text{em}\,,
\end{align}
where $\vbl$ is the intermediate symmetry breaking scale and $v_\text{EW}$ is the electroweak symmetry breaking scale. For example, the first stage of the above symmetry breaking chain can be achieved by $SU(2)_R$ triplet of vanishing $B-L$ charge while the second stage can be taken care of by a triplet of non-zero $B-L$ charge. In such a scenario, the bias term can be written as dimension five operator involving both types of triplet scalars. Following \cite{Borah:2022wdy}, one can show that the bias term is related to the two scales of symmetry breaking via $\Delta V\simeq u^3\,\vbl^2/M_P$. In that case, one finds, $\Tann\propto \vbl$ [cf.Eq.~\eqref{eq:Tann}]. We show constraint on the LR symmetry breaking scale $u$, that allows to be within 2-$\sigma$ of the NANOGrav 15 yr data in Fig.~\ref{fig:fig5}. Corresponding to each $\Tann$, the $\vbl$ is fixed, as mentioned in the upper axis label. We find, in order to be compatible with 1-$\sigma$ contour of the NANOGrav 15 yr data, $0.04 \, {\rm GeV} \lesssim\vbl\lesssim 1$ GeV, while $ 10^2 \, {\rm TeV} \lesssim u\lesssim 10^3$ TeV. However, such low scale $\vbl$ will lead to light $Z'$ gauge bosons, already ruled out by the large hadron collider (LHC) data \cite{ATLAS:2019erb, CMS:2018ipm}.

\vspace{0.5cm}
\noindent

{\bf Conclusion:} We have investigated the consequence of the recent PTA results on the scale of Dirac leptogenesis and left-right symmetric model. In minimal version of both these scenarios, domain walls arise due to spontaneous breaking of a discrete $Z_2$ symmetry. While the bound on the scale of Dirac leptogenesis from PTA data depend upon the size of Dirac Yukawa couplings, for conservative choice of such couplings with fine-tuning at the level of the SM, we find a lower bound $M_N > 10^7$ GeV, keeping leptogenesis at intermediate scales. However, for order one Yukawa coupling, this bound is much stronger $M_N > 10^{17}$ GeV keeping only non-thermal Dirac leptogenesis option viable [cf. Fig.~\ref{fig:fig3}]. Due to the constrained structure of the minimal LRSM, we get much tighter constraint on the scale of left-right breaking namely $M_{\rm LR} \sim 10^6$ GeV [cf. Fig.~\ref{fig:fig4}], in order to satisfy NANOGrav 15 yr data, keeping the model out of reach from direct search experiments like the LHC. For non-minimal LRSM, we can fit NANOGrav 15 yr data for $M_{\rm LR}\simeq \{120-1050\}$ TeV [cf. Fig.~\ref{fig:fig5}] but with a very low $B-L$ breaking scale ruled out by the LHC data. Future data from PTA or other GW experiments are expected to shed more light on the parameter space of this model, by constraining the spectrum at higher frequencies.

\acknowledgements
The work of D.B. is supported by the science and engineering research board (SERB), Government of India grant MTR/2022/000575. 

\twocolumngrid
\bibliographystyle{apsrev}
\bibliography{ref.bib, ref1.bib,ref2.bib,ref3.bib,ref4.bib}

\end{document}